\documentclass[a4paper,aps,prx,reprint,floatfix,amsmath,amssymb,amsfonts,longbibliography,superscriptaddress]{revtex4-2}
\usepackage{mathptmx}
\usepackage[scaled=0.9]{helvet}
\usepackage[utf8]{inputenc}
\usepackage{textcomp}
\usepackage{amsmath}
\usepackage{graphicx}
\usepackage{esint}
\usepackage[pdftex]{hyperref}

\hypersetup{pdfauthor={Vladimir Kaganer}, bookmarksnumbered=true, pdftitle={X-ray scattering study of GaN nanowires grown on Ti/Al$_{2}$O$_{3}$ by molecular beam epitaxy}, colorlinks, citecolor=blue, linkcolor=blue, urlcolor=blue}

\makeatletter

\pdfpageheight\paperheight
\pdfpagewidth\paperwidth

\providecommand{\tabularnewline}{\\}

\usepackage{upgreek}
\usepackage[dvipsnames]{xcolor}
\usepackage{soul}
\usepackage{xspace}
\usepackage[rightcaption]{sidecap}
\usepackage{epstopdf}
\usepackage{units}

\DeclareMathAlphabet{\mathcal}{OMS}{cmsy}{m}{n} 

\setcitestyle{numbers,square}

\makeatother

\begin{document}
\title{X-ray scattering study of GaN nanowires grown on Ti/Al$_{2}$O$_{3}$
by molecular beam epitaxy}
\author{Vladimir M. Kaganer}
\affiliation{Paul-Drude-Institut für Festkörperelektronik, Leibniz-Institut im
Forschungsverbund Berlin e.~V., Hausvogteiplatz 5--7, 10117 Berlin,
Germany}
\author{Oleg V. Konovalov}
\affiliation{ESRF -- The European Synchrotron, 71 avenue des Martyrs, 38043 Grenoble,
France}
\author{Gabriele Calabrese}
\affiliation{Paul-Drude-Institut für Festkörperelektronik, Leibniz-Institut im
Forschungsverbund Berlin e.~V., Hausvogteiplatz 5--7, 10117 Berlin,
Germany}
\affiliation{Istituto per la Microelettronica e Microsistemi, Consiglio Nazionale
delle Ricerche, via Gobetti 101, 40129 Bologna, Italy}
\author{David van Treeck}
\affiliation{Paul-Drude-Institut für Festkörperelektronik, Leibniz-Institut im
Forschungsverbund Berlin e.~V., Hausvogteiplatz 5--7, 10117 Berlin,
Germany}
\author{Albert Kwasniewski}
\author{Carsten Richter}
\affiliation{Leibniz-Institut für Kristallzüchtung (IKZ), Max-Born-Str.\ 2, 12489
Berlin, Germany}
\author{Sergio Fernández-Garrido}
\affiliation{Paul-Drude-Institut für Festkörperelektronik, Leibniz-Institut im
Forschungsverbund Berlin e.~V., Hausvogteiplatz 5--7, 10117 Berlin,
Germany}
\affiliation{Institute for Optoelectronic Systems and Microtechnology (ISOM), Universidad Politécnica de Madrid, Avda.\ Complutense 30, 28040 Madrid, Spain}
\author{Oliver Brandt}
\affiliation{Paul-Drude-Institut für Festkörperelektronik, Leibniz-Institut im
Forschungsverbund Berlin e.~V., Hausvogteiplatz 5--7, 10117 Berlin,
Germany}
\date{\today}
\begin{abstract}
GaN nanowires (NWs) grown by molecular beam epitaxy on Ti films sputtered
on Al$_{2}$O$_{3}$ are studied by X-ray diffraction (XRD) and grazing
incidence small-angle X-ray scattering (GISAXS). XRD, performed both
in symmetric Bragg reflection and at grazing incidence, reveals Ti,
Ti$_{3}$O, Ti$_{3}$Al, and TiO$_x$N$_y$ crystallites with in-plane and out-of-plane
lattice parameters intermediate between those of Al$_{2}$O$_{3}$
and GaN. These topotaxial crystallites in Ti film, formed due to interfacial reactions 
and N exposure, possess fairly little misorientation with respect to  Al$_{2}$O$_{3}$.
As a result, GaN NWs grow on the top TiN layer possessing a high degree of 
epitaxial orientation with respect to the substrate. The measured GISAXS intensity
distributions are modeled by the Monte Carlo method taking into
account the orientational distributions of NWs, a variety of their
cross-sectional shapes and sizes, and roughness of their side facets.
The cross-sectional size distributions of the NWs and the relative
fractions of $(1\bar{1}00)$ and $(11\bar{2}0)$ side facets are determined. 
\end{abstract}
\maketitle

\section{Introduction}

Semiconductor nanowires (NWs) have essential advantages over epitaxial
films of the same materials due to large areas of their side facets
as well as the ability of free elastic relaxation of the material,
which provides both a reduction of the density of lattice defects
near the interface to the substrate and defect free interfaces in
axial and radial NW heterostructures. The self-induced growth of GaN
NWs \cite{garrido09,geelhaar11} does not involve, in contrast to
the vapour--liquid--solid growth of the majority of semiconductor
materials, metal particles at the top \cite{ristic08}. GaN NWs grow
on various substrates in dense arrays and, as a consequence of the
self-induced growth, their density can hardly be controlled by varying
temperature or the atomic fluxes. NWs in dense arrays shadow the side
facets of each other from the impinging fluxes \cite{sibirev12,sabelfeld13},
hindering the growth of radial heterostructures, and also bundle together
\cite{kaganer16bundling}. TiN has been found to be a substrate with
a low nucleation rate of GaN NWs, resulting in a density an order
of magnitude lower than that of GaN NWs grown on Si(111). TiN layers
have been prepared by nitridation of Ti films \cite{sarwar15,woelz15,zhao16,treeck18,mudiyanselage20}
and Ti foils \cite{calabrese16,may16,calabrese17,ramesh19,ramesh20,mudiyanselage20},
as well as by directly sputtering TiN$_{x}$ on Al$_{2}$O$_{3}$
\cite{auzelle21}.

Recently, we have applied grazing incidence small-angle X-ray scattering
(GISAXS) to study dense arrays of GaN NWs on Si(111) \cite{kaganer21gisaxs}.
We have shown that GISAXS is well suited to obtain statistical information
on the average radius and the width of the radii distribution of a
NW array. GISAXS is also sensitive to the roughness of the side facets
of NWs, which has been found to be less than 1~nm, i.e., 3--4 times
the height of the atomic steps. We have shown that the epitaxial orientation
of the NWs gives rise to a dependence of the GISAXS intensity on the
sample orientation with respect to the X-ray beam. The intensity is
maximum along the normals to the side facets, which is due to facet
truncation rod scattering, similar to the crystal truncation rods
from planar crystals. The approach has been developed initially for
NWs represented by prisms with hexagonal cross sections \cite{kaganer21gisaxs},
i.e., with the GaN$(1\bar{1}00)$ side facets, and then also applied
to NWs with both $(1\bar{1}00)$ and $(11\bar{2}0)$ side facets to
determine the ratio of the areas of these two facets \cite{volkov22}.

In the present paper, we apply GISAXS to study GaN NWs on nitridated
Ti films sputtered on Al$_{2}$O$_{3}(0001)$. We develop further
the approach proposed for the analysis of dense arrays of GaN NWs
on Si(111) \cite{kaganer21gisaxs}. The GISAXS intensity distribution
contains a weaker intensity from a lower NW density which overlaps
with a stronger parasitic signal from the sputtered film, which makes
the analysis more complicated. These two contributions can be distinguished
in the intensity pattern and the NW intensity can be extracted since
NWs are needle-shaped oriented objects whose intensity in reciprocal
space resembles a disk perpendicular to the long axis of the NWs.
The intensity distributions from NWs are modeled by the Monte Carlo
method that takes into account the distributions of the NW shapes
and orientations. By comparing the measured and the simulated intensities,
we find the distribution of the NW radii and that of the ratio of
the $(1\bar{1}00)$ and $(11\bar{2}0)$ side facets, as well as the
roughness of these facets.

The Monte Carlo modeling of the GISAXS intensity requires as an input
the range of orientations of the NW long axes (tilt) and that of the
side facets (twist). As a prerequisite of the GISAXS study, we perform
X-ray diffraction (XRD) measurements with the primary purpose to determine
these ranges from the widths of the respective reflections on sample
rotation. We find that the arrays of GaN NWs on Ti/Al$_{2}$O$_{3}$
possess notably smaller tilt and twist ranges than their counterparts
on Si(111). The XRD measurements reveal also the crystalline phases
formed in the sputtered Ti film on chemical reactions with the substrate
material, which sheds light on the surprisingly narrow orientational
distributions of the NWs. Hence, we present below the results of XRD
in some detail prior to presenting the GISAXS results.

\section{Experiment}
\label{sec:Experiment}

For the present study, we have chosen four samples identical to samples
A--D in Ref.~\cite{calabrese20}, and keep the same notation of
the samples. The growth conditions and the results of the XRD and
GISAXS studies of the present work are summarized in Table~\ref{Table}. 

\begin{table*}
\begin{tabular}{|c||c|c|c|c|c|c||c||c|c||c|c|c|c|c|}
\hline 
sample  & \multicolumn{6}{c||}{growth} & {SEM} & \multicolumn{2}{c||}{XRD} & \multicolumn{5}{c|}{GISAXS}\tabularnewline
\hline 
 & $\Phi_{\mathrm{Ga}}$  & $\Phi_{\mathrm{N}}$  & $T$  & $t$  & $T_{\mathrm{Ti}}$  & $L_{\mathrm{NW}}$& radius  & tilt  & twist  & $\alpha_{i}$  & tilt  & $\sigma$  & radius  & $f_{11\bar{2}0}$\tabularnewline
\hline 
 & \multicolumn{2}{c|}{ML/s} & $^{\circ}$C  & min  & \textmu m  & \textmu m  & nm& \multicolumn{2}{c||}{deg.} & deg.  & deg.  & nm  & nm  & \tabularnewline
\hline 
A  & 0.27  & 0.36  & 710  & 240  & 1.3  & 1.6 & 22 & 1.86  & 0.73  & 0.25  & 2.3  & 2.3  & 20.7$\pm$8.6  & 0.4$\pm$0.09\tabularnewline
\hline 
B & 0.32  & 0.75  & 630  & 120  & 3.4  & 1.0 & 15 & 1.78  & 0.99  & 0.36  & 1.7  & 0.7  & 11.2$\pm$5.2  & 0\tabularnewline
\hline 
C & 0.39  & 1.05  & 610  & 60  & 3.4  & 0.7 & 22 & 1.52  & 0.89  & 0.6  & 1.7  & 1.3  & 17.4$\pm$8.6  & 0.33$\pm$0.14\tabularnewline
\hline 
D & 0.32  & 0.75  & 600  & 120  & 3.4  & 1.2 & 29 & 1.52  & 0.82  & 0.29  & 2.0  & 0.9  & 24.4$\pm$10.3  & 0.37$\pm$0.14\tabularnewline
\hline 
\end{tabular}\caption{Growth parameters of the NW ensembles, their average radii estimated from top-view SEM micrographs \cite{calabrese20}, and results of the present XRD and GISAXS study. The growth parameters are Ga and N fluxes $\Phi_{\mathrm{Ga}}$
and $\Phi_{\mathrm{N}},$ growth temperature $T$, growth time $t$,
thickness of the sputtered Ti layer $T_{\mathrm{Ti}}$, and the NW
length $L_{\mathrm{NW}}.$ The FWHM of the tilt and twist distributions
are measured by XRD in $0002$ and $2\bar{2}00$ reflections, respectively.
The incidence angle $\alpha_{i}$ in the GISAXS measurements is chosen
for each sample to provide the best visibility of the NW intensity.
The FWHM of the tilt are obtained from the Monte Carlo simulation
of the $q_{z}$ scans in Fig.~\ref{fig:tilt}, while the roughness
$\sigma$ of the side facets, the NW radius and the fraction $f_{11\bar{2}0}=p_{11\bar{2}0}/(p_{11\bar{2}0}+p_{1\bar{1}00})$
of the $\{11\bar{2}0\}$ facets are obtained from the Monte Carlo
simulations of the $I(q_{x})$ intensity curves in Fig.~\ref{fig:Iq4}.
The two numbers for the radius and facet ratios are the mean value
and the standard deviation.}
\label{Table} 
\end{table*}

The samples are grown by plasma-assisted molecular beam epitaxy (PA-MBE)
on Ti films sputtered on Al$_{2}$O$_{3}$(0001). Before NW growth,
a Ti film with a thickness of either 1.3~\textmu m (sample A) or
3.4~\textmu m (samples B--D) is deposited on bare Al$_{2}$O$_{3}$(0001)
by magnetron sputtering as described elsewhere \cite{treeck18}.
After Ti sputtering, the samples are loaded into the growth chamber
of the PA-MBE system, being exposed to air during the transfer. The MBE
system is equipped with a solid-source effusion cell for Ga and a
radio-frequency N$_{2}$ plasma source for active N. The impinging
Ga and N fluxes are expressed in monolayers per second (ML/s). The
substrate temperature during NW growth is measured with a thermocouple
placed in contact with the substrate heater. For samples B and C,
a dedicated nitridation step is introduced before NW growth. The N
flux used for substrate nitridation is the same one as for GaN growth.
After the intentional substrate nitridation process, the Ga shutter
is opened to initiate the formation of GaN NWs. For samples A and
D, the Ti film is nitridated after opening simultaneously the Ga and
N shutters to initiate the growth of GaN. Further details concerning
the nitridation process can be found in Ref.~\cite{calabrese19}.

\begin{figure}
\includegraphics[width=1\columnwidth]{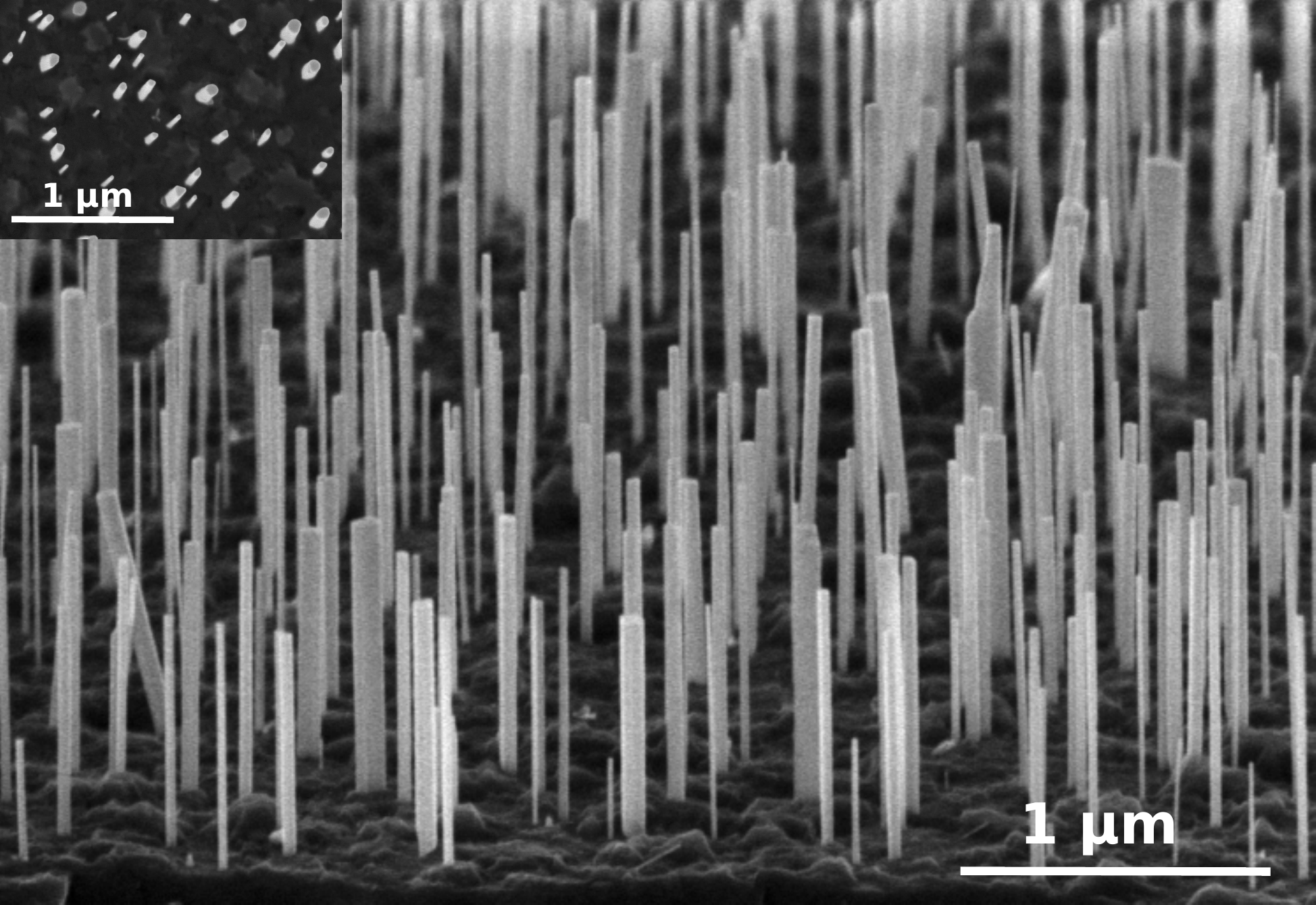}
\caption{Bird's-eye view scanning electron micrograph of sample C. The inset shows a scanning electron micrograph of the same sample taken in a direction close to the surface normal. }
\label{fig:SEM} 
\end{figure}

Figure \ref{fig:SEM}
presents a scanning electron microscopy (SEM) micrograph of sample C. A good alignment of the NWs in the direction
of the substrate surface normal is clearly seen in the figure and quantified below
using XRD. A large roughness of the substrate surface is also evident from the figure, giving rise to an additional scattering in the GISAXS experiment described below. The
inset in Fig.~\ref{fig:SEM} shows a scanning electron micrograph of the same sample
taken in the direction close to the surface normal. Such images were used to obtain
the average NW radii \cite{calabrese20} as included in 
Table~\ref{Table}. However, the resolution of the SEM micrographs is
not good enough to reveal details of the cross-sectional shapes of NWs, that we find in the GISAXS study below.

The laboratory XRD measurements are performed in two geometries, out-of-plane
in the familiar symmetric Bragg reflection and in-plane at grazing
incidence. For the acquisition of out-of-plane reciprocal space maps,
we use a Bruker D8 Discover Diffractometer operating a Cu anode at
1.6~kW. The beam conditioning optics are comprised of a Göbel mirror
and an asymmetric 2-bounce channel cut Ge (220) monochromator. The
beam is thus collimated to about 0.0085$^{\circ}$. The diffracted
intensity is recorded using a Mythen 1D position sensitive detector
with a channel pitch and width of 50 \textmu m and 8 mm, respectively.
The two-dimensional reciprocal space maps are obtained by rocking
scans of the sample.

The in-plane XRD measurements are performed using a Rigaku SmartLab
high-resolution X-ray diffractometer in a horizontal scattering plane.
The source is a 0.4$\times$8 mm$^{2}$ electron line focus on a rotating
Cu anode operating at 9 kW. The beam is vertically collimated by a
Göbel mirror and an asymmetrically cut 2-bounce Ge (220) monochromator
is used to select the Cu K$\alpha_{1}$ emission line. The beam size
is reduced to 0.3$\times$5 mm$^{2}$ by slits and the angular resolution
is defined by horizontal Soller slits of 0.25$^{\circ}$ and 0.228$^{\circ}$
angular acceptance for the incident and scattered beams, respectively.
The incidence angle is set to 0.25$^{\circ}$. Diffraction intensities
are recorded using a Hypix-3000 area detector with 100$\times$100
\textmu m$^{2}$ pixel size.

Grazing incidence X-ray diffraction (GID) reciprocal space maps are
aquired at the beamline ID10 of the European Synchrotron Radiation
Facility (ESRF) at an X-ray energy of 22 keV (wavelength $\lambda=0.5636$~Å).
A linear detector (Mythen 1K, Dectris) is placed parallel to the substrate
surface to cover the range of scattering angles around the diffraction
angle of the GaN(1$\bar{1}$00) reflection with $2\theta\approx11.7^{\circ}$.
The grazing incidence angle is $0.12^{\circ}$ and the exit angle
is $0.24^{\circ}$. The sample is rotated about the substrate surface
normal, and linear detector scans are recorded for different azimuthal
angles $\omega$. The obtained reciprocal space maps are similar to
the $\omega-2\theta$ maps measured in laboratory XRD experiments.
We refer to them as the $\omega-2\theta$ maps, although the sample
rotation axis is along the surface normal, in contrast to symmetric
Bragg reflections in the laboratory XRD measurements where the sample
rotation axis lies in the surface plane.

GISAXS measurements are also performed at the beamline ID10 of ESRF
at the same X-ray energy of 22~keV. The incident beam is directed
at grazing incidence to the substrate. The grazing angle $\alpha_{i}$,
presented in Table~\ref{Table}, is chosen for each sample from several
trials to provide the best signal from NWs. The critical angle of
the total external reflection for Al$_{2}$O$_{3}$ at the used energy
is $0.1^{\circ}$. Hence, the angles $\alpha_{i}$ are at least 2.5
times larger than the critical angle of the substrate, which allows
us to avoid possible complications of the scattering pattern typical
for grazing incidence X-ray scattering \cite{renaud09}. 
The X-ray beam incident on the sample is  focused by compound refractive lenses to the 
size at the sample position of 135~µm laterally and 13~µm vertically
(in the direction of the surface normal). At an incidence angle of 0.25$^\circ$, the
size of the spot illuminated by the incident beam at the sample is $3\times0.135$~mm$^2$.
With a typical NW density of $1\times10^9$~cm$^{-2}$, approximately $4\times10^6$ NWs are
illuminated simultaneously. A two-dimensional
detector (Pilatus 300K, Dectris) is placed at a distance of 2.38~m
from the sample. The angular width of a detector pixel is $8.06\times10^{-3}$~nm$^{-1}$. 

\section{Results}

\subsection{XRD}

\label{subsec:XRD}

\begin{figure*}
\includegraphics[width=1\textwidth]{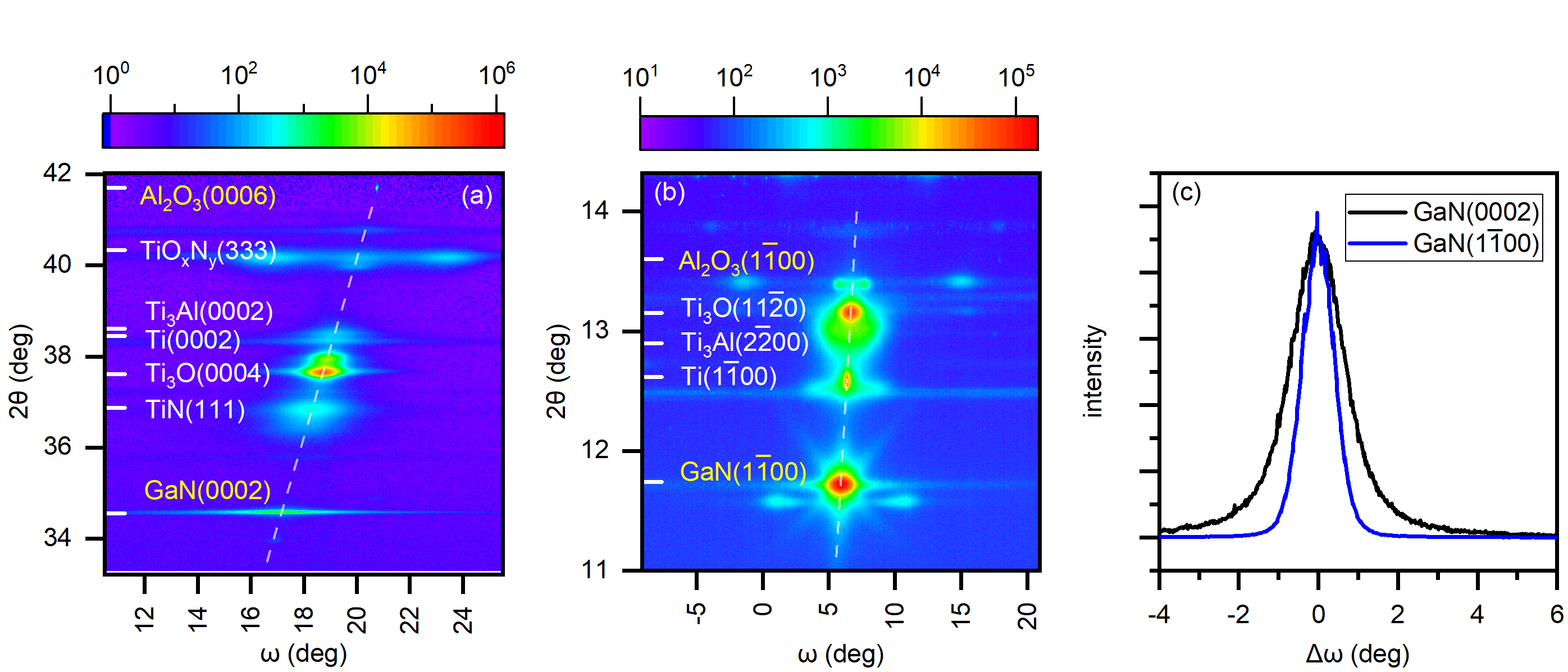}

\caption{(a) XRD reciprocal space map in the vicinity of the GaN($0002$) reflection,
(b) GID reciprocal space map in the vicinity of the GaN($1\bar{1}00$)
reflection and (c) intensity profiles through the GaN reflections
from sample C. The color-coded scale bars represent the intensity
in counts, while the dashed lines show the $\omega=2\theta$ radial
scans across the Al$_{2}$O$_{3}$ substrate and the GaN NW reflections.
The peaks along these lines in between the substrate and the NW reflections
are due to different crystalline phases emerging due to interfacial
reactions in the sputtered Ti film on the substrate. Plausible crystal
structures are indicated.}

\label{fig:XRD} 
\end{figure*}

Laboratory XRD measurements of the $\omega-2\theta$ reciprocal space
maps at the GaN(0002) reflection are performed to measure the NW tilt.
The maps for samples A--D are very similar. Figure \ref{fig:XRD}(a)
presents such a map for sample C. One can see a very sharp Al$_{2}$O$_{3}$(0006)
reflection and a GaN(0002) reflection that is extended in the $\omega$
direction. An $\omega$ scan through the GaN(0002) reflection, extracted
from the map, is shown in Fig.~\ref{fig:XRD}(c), yielding the NW
tilt, i.\,e., the FWHM of the NW out-of-plane orientation distribution
of 1.52$^{\circ}$. Similar measurements on the other samples give
close results summarized in Table \ref{Table}.

The dashed line from the Al$_{2}$O$_{3}$(0006) to GaN(0002) reflections
in Fig.~\ref{fig:XRD}(a) indicates the direction of a radial $\omega-2\theta$
scan. One can see a number of peaks in between. They are the result
of the interfacial reactions in the sputtered Ti film \cite{selverian91,li92,koyama93,kelkar95,calabrese19}.
High-resolution transmission electron microscopy has shown that the
top 40--80~nm of the Ti film is transformed to TiN by nitridation,
so that GaN NWs grow on TiN \cite{calabrese19}.

We indicate in Fig.~\ref{fig:XRD}(a) the $2\theta$ angles for plausible
compounds and reflections found in Pearson's Crystal Data \cite{pearson}.
We searched for hexagonal phases, with the basal plane parallel to
the substrate surface, of compounds composed of Ti and the chemical
elements of the substrate or the NWs. Since positions of the diffraction
peaks are affected by thermal strain, epitaxial strain, and also by
impurities in the crystals, we do not expect that the $2\theta$ values
exactly coincide with the literature data. Also, there is a notable
scattering in the lattice parameters between results of different
studies collected in the database \cite{pearson}. We also include
(111) oriented TiO$_{x}$N$_{y}$ oxynitrides with rocksalt structure, 
revealed by X-ray photoelectron
spectroscopy in previous studies \cite{milosev95,cheng07,calabrese17}, and with a lattice parameter only weakly depending on $x$ and $y$ 
\cite{pearson}. The strongest
reflection in Fig.~\ref{fig:XRD}(a) is Ti$_{3}$O(0004). It possesses
a width in $\omega$ direction of 0.5$^{\circ}$, i.e., the tilt of
the Ti$_{3}$O crystallites is three times smaller compared to that
of the GaN NWs.

Figure \ref{fig:XRD}(b) shows the GID $\omega-2\theta$ map in the
vicinity of the GaN($1\bar{1}00$) reflection of the same sample C.
Upon rotation of the sample about the surface normal, the diffraction
pattern of Fig.~\ref{fig:XRD}(b) is repeated after every 60$^{^{\circ}}$\cite{calabrese20}.
The dashed line represents the $\omega-2\theta$ scan. The Al$_{2}$O$_{3}$($1\bar{1}00$)
substrate reflection is not seen because the small incidence and exit
angles prevent the penetration of the x-ray radiation into the substrate.
This reflection is observed in a measurement with larger incidence
and exit angles (not shown here). The scattering angles $2\theta$
of the same phases as in Fig.~\ref{fig:XRD}(a) are indicated. We
find the in-plane reflections of the hexagonal crystals Ti, Ti$_{3}$O,
and Ti$_{3}$Al. The cubic TiN does not have a Bragg reflection in
the angular range of the map in Fig.~\ref{fig:XRD}(b).

The $\omega$-scan through the GaN($1\bar{1}00$) reflection is presented
in Fig.~\ref{fig:XRD}(c). Its FWHM yields the NW twist, i.\,e.,
the NW in-plane orientation distribution width, of 0.96$^{\circ}$.
For a comparison, the measurement of the same reflection by triple-crystal
diffraction at the laboratory diffractometer gives a somewhat smaller
value of 0.89$^{\circ}$. The difference can be explained by the geometrical
broadening in the synchrotron measurement with the linear detector.
In the latter measurement, an $\approx5$~mm long stripe on the sample
surface is illuminated by the incident X-ray beam and contributes
to diffraction. Hence, the twist values obtained by the laboratory
X-ray diffraction are more reliable and are given in Table~\ref{Table}.

\subsection{GISAXS}

\begin{figure*}
\includegraphics[width=1\textwidth]{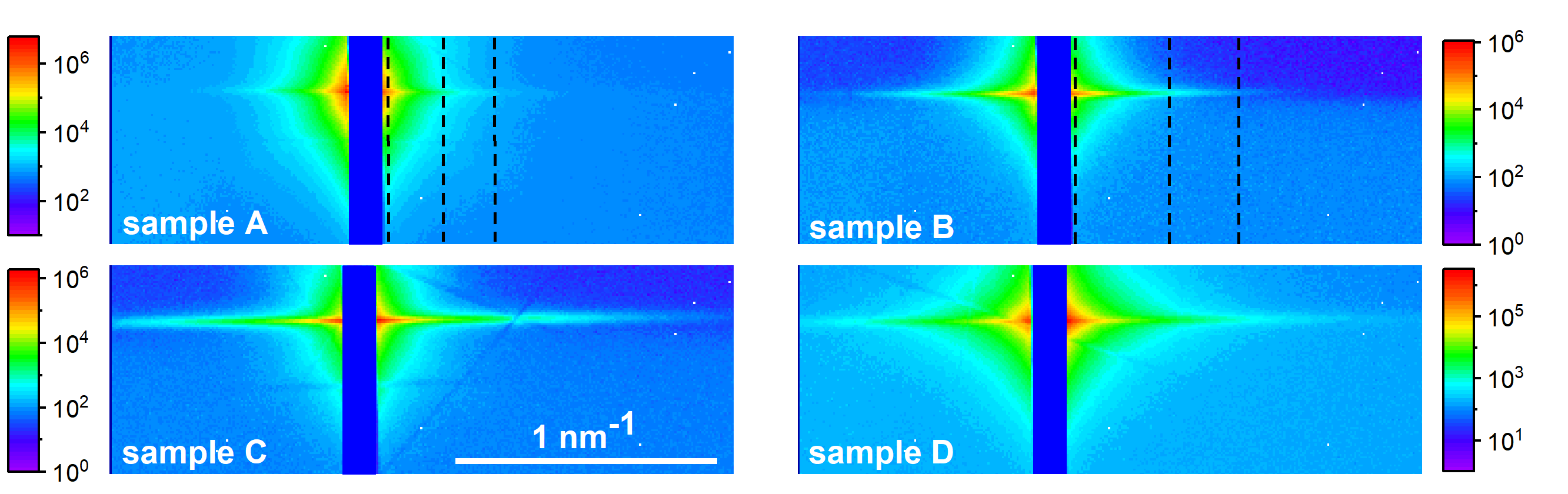}

\caption{GISAXS intensity from samples A--D as measured by a two-dimensional
detector. The scattering around the transmitted beam is shown. The
vertical blue bar in the middle of each scattering patterns is the
beamstop. The three vertical dashed lines at the maps of samples A
and B mark the positions of the scans presented in Fig.~\ref{fig:scans}.
The color-coded scale bars represent the intensity in counts. The
incidence angles $\alpha_{i}$ of the X-ray beam with respect to the
substrate are given in Table \ref{Table}.}

\label{fig:maps} 
\end{figure*}

Figure \ref{fig:maps} presents the GISAXS intensity distributions
around the transmitted beam for samples A--D. Since the GISAXS experiment
is intentionally performed with the incidence angles $\alpha_{i}$
exceeding at least 2.5 times the critical angle for the substrate
(see Table~\ref{Table}), the small-angle X-ray scattering patterns
comprise three separate regions: around the transmitted beam, around
the beam reflected from the substrate, and the Yoneda streak (see
Fig.~2 in Ref.~\cite{kaganer21gisaxs}). We choose for the analysis
the scattering around the transmitted beam, since it is more intense.
One can see on each map in Fig.~\ref{fig:maps} a horizontal (parallel
to the substrate surface and normal to the long NW axis) streak and
a halo around the direct beam direction. The streak is due to scattering
from NWs: the NWs are long rods that scatter in the plane perpendicular
to the rods. The halo stems from the scattering from the sputtered
film on the substrate.

\begin{figure}[b]
\includegraphics[width=1\columnwidth]{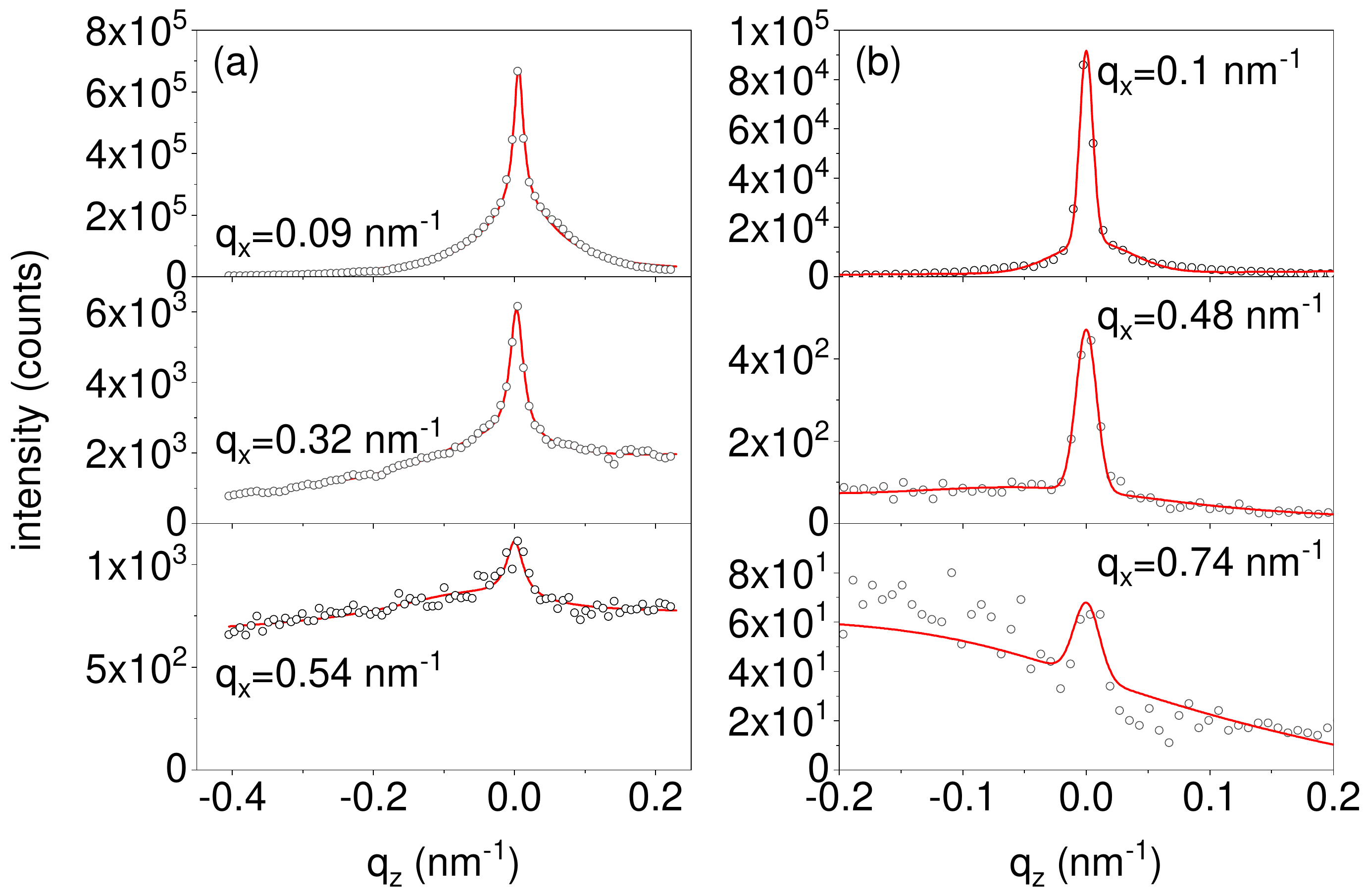}

\caption{Measured intensity profiles (circles) along lines of constant $q_{x}$
marked by the dashed lines in Fig.~\ref{fig:maps}, and the respective
fits to a sum of two Gaussians plus a background (lines) for samples
(a) A and (b) B.}

\label{fig:scans} 
\end{figure}

\begin{figure}[b]
\includegraphics[width=1\columnwidth]{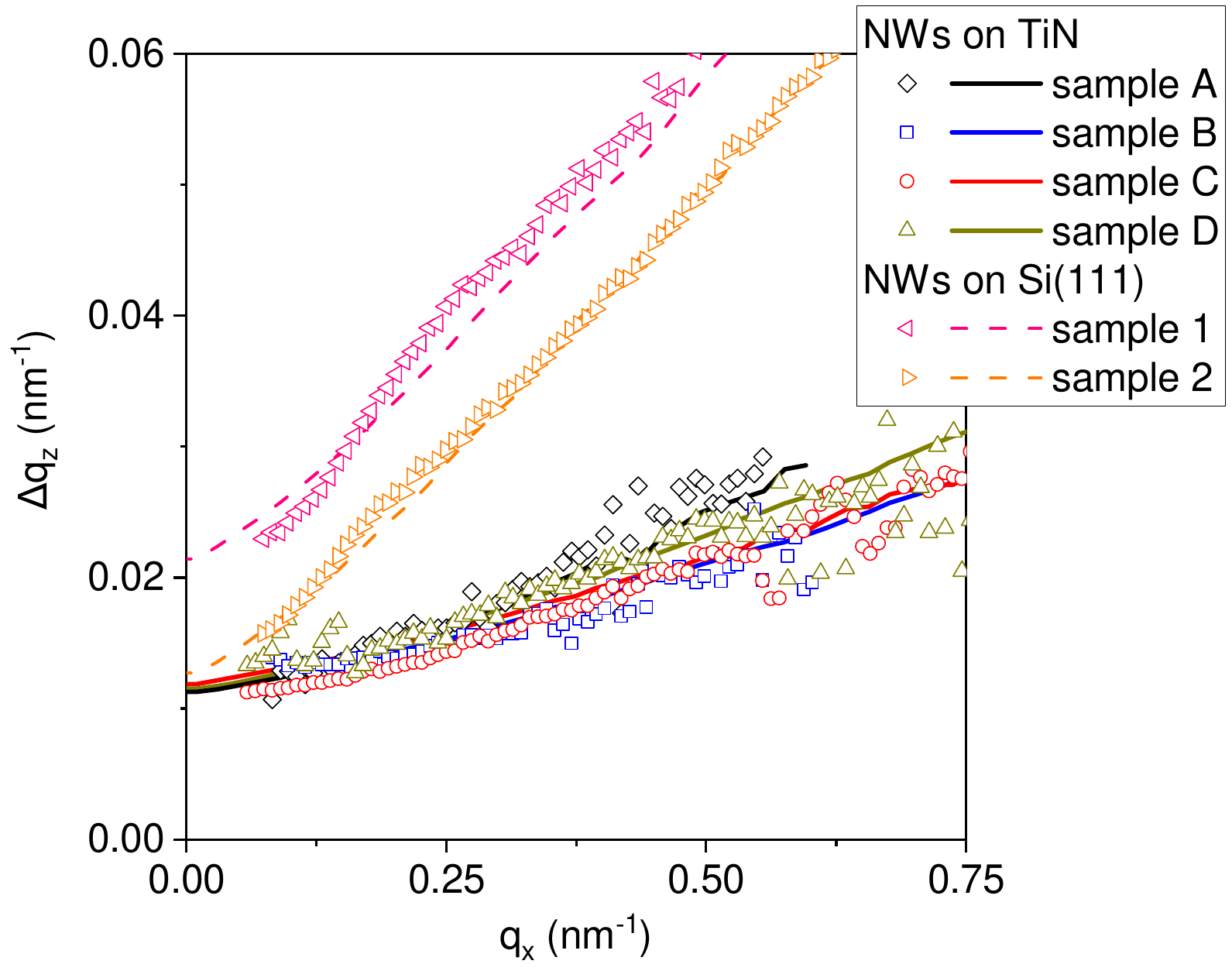}

\caption{FWHM of the intensity profiles $\Delta q_{z}$ as a function of the
wavevector $q_{x}$ (open symbols). The experimental data for samples
A--D are compared with the respective data for GaN NW ensembles on
Si(111) (samples 1 and 2 in Ref.~\cite{kaganer21gisaxs}). Lines
are the results of the Monte Carlo simulations with the ranges of
the NW tilt angles presented in Table \ref{Table} and a resolution
of $1.2\times10^{-2}$~nm$^{-1}$.}

\label{fig:tilt} 
\end{figure}

These two contributions to the scattered intensity can be distinguished
by analyzing the scans in the surface normal direction. Some exemplary
scans are marked in Fig.~\ref{fig:maps} by dashed lines and the
intensity in these scans is shown in Fig.~\ref{fig:scans}. Here,
$q_{x}$ and $q_{z}$ are parallel to the substrate surface and normal
to it, respectively. The scans are fitted to a sum of two Gaussians,
a narrow one representing the X-ray scattering intensity from the
NWs and a broad one due to the X-ray scattering from the sputtered
film, plus a background that linearly depends on $q_{z}$. Since each
NW scatters in the plane perpendicular to its long axis, a range of
orientations of the long axes (tilt) gives rise to a fan in the intensity
distribution. Hence, the full width at half maximum (FWHM) $\Delta q_{z}$
of a $q_{z}$-scan of the intensity is expected to depend linearly
on $q_{x}$. The $\Delta q_{z}(q_{x})$ dependencies obtained from
the fits of the $q_{z}$-scans for samples A--D are presented in
Fig.~\ref{fig:tilt}. The expected linear dependence is observed.
Figure \ref{fig:tilt} also contains the respective data for GaN NWs
on Si(111) \cite{kaganer21gisaxs}.

The lines in Fig.~\ref{fig:tilt} are the result of a Monte Carlo
simulation that takes into account the distributions of the NW cross-sectional
sizes, lengths, and orientations \cite{kaganer21gisaxs}. We simulate
the $q_{z}$ dependence of the scattered intensity $I(q_{z})$ at
a given $q_{x}$ and then fit this curve to a Gaussian, in the same
way as it is done with the experimental data. The $q_{x}$ dependence
of the FWHM $\Delta q_{z}$ of the Gaussians is close to straight
lines, whose slopes depend on the width of the distribution of the
NW tilt angles. The tilt angles obtained by adjusting the simulated
curves to the experimental ones are presented in Table \ref{Table}.

The width $\Delta q_{z}$ at $q_{x}=0$ has been treated in Ref.~\cite{kaganer21gisaxs}
as a broadening solely due to the finite NW lengths $L$, $\Delta q_{z}\propto L^{-1}$.
The lengths thus obtained were found to be smaller than the actual
NW lengths and attributed to the lengths of the segments of bundled
NWs between the joints. In the present case of low NW density, the
NWs are not bundled, and their lengths, obtained from the side view
SEM micrographs and given in Table \ref{Table},
are used as an input in the Monte Carlo simulation. We find in the
simulation, that the finite-length broadening $\Delta q_{z}$ at $q_{x}=0$
is notably smaller than the widths found in the experiment. Then,
we take into consideration a finite resolution of the experimental
curves and accordingly perform an average of the Monte Carlo intensities
$I(q_{z})$ in a range $\Delta q^{\mathrm{res}}$. The curves presented
in Fig.~\ref{fig:tilt} are obtained with the resolution $\Delta q^{\mathrm{res}}=1.2\times10^{-2}$~nm$^{-1}$,
which is 1.5 times the angular size of the detector pixel. This result
can be considered as a partial exposure of the neighbor detector pixels.
Thus, the curves in Fig.~\ref{fig:tilt} are obtained taking into
account both the finite-length and the resolution broadening of the
Monte Carlo simulated curves.

We reproduce in Fig.~\ref{fig:tilt} also the experimental data for
the GISAXS intensity from GaN NWs on Si(111) \cite{kaganer21gisaxs}
and perform new Monte Carlo simulations, which now include the NW
lengths taken from the SEM micrographs and the resolution $\Delta q^{\mathrm{res}}$
obtained above. The width $\Delta q_{z}$ at $q_{x}=0$ for the 230~nm
long NWs in sample 1 is mainly due to finite-length broadening, while
the respective width for the 650~nm long NWs in sample 2 is mainly
due to the finite resolution.

It is evident from comparison of the slopes of the curves in Fig.~\ref{fig:tilt},
that the NWs on Ti/Al$_{2}$O$_{3}$ show a notably narrower orientation distribution compared
to the NWs on Si(111). The widths of the tilt angle distributions
obtained from Fig.~\ref{fig:tilt} are found to range from 1.7$^{\circ}$
to 2.3$^{\circ}$ for samples A--D (see Table \ref{Table}), while
for NWs on Si(111), they amount to 5.1$^{\circ}$ and 4.9$^{\circ}$
for samples 1 and 2, respectively.

For further analysis, we need the GISAXS intensity $I(q_{x})$ along
the horizontal stripes in the reciprocal space maps in Fig.~\ref{fig:maps}.
These stripes correspond to the maximum intensity of the $q_{z}$-scans
exemplified in Fig.~\ref{fig:scans}, with the background scattering
subtracted. We use the established linear dependence of the widths
$\Delta q_{z}$ on $q_{x}$ in Fig.~\ref{fig:tilt} to improve the
fits of the $q_{z}$-scans and extend the $q_{x}$-range as much as
possible to the regions of low intensity and substantial noise in
the experimental data (see the bottom curves in Fig.~\ref{fig:scans}).
After a fit of the $q_{z}$-scans from a reciprocal space map is performed
and the widths $\Delta q_{z}$ obtained, the dependence $\Delta q_{z}(q_{x})$,
shown in Fig.~\ref{fig:tilt}, is fitted by a straight line. Then,
the fit of the $q_{z}$-scans is repeated, now with the peak position
and width thus predefined, rather than to be free parameters of the
fit. Since the only output of this second fit that we need is the
maximum intensity, it can be determined at large $q_{x}$, where the
experimental data are noisy. The curves shown in Fig.~\ref{fig:scans}
are the result of such two-step fits.

The GISAXS intensity $I(q_{x})$ is expected to follow Porod's law
$I(q_{x})\propto q_{x}^{-4}$ at large $q_{x}$ , which describes
small--angle scattering from any particle with a sharp change of
the electron density at the surface \cite{porod51}. It has the same
nature and is as general as Fresnel's law for scattering from planar
surfaces \cite{sinha88}. The epitaxy of GaN NWs to the substrate
gives rise to preferential orientations of the side facets and results
in an azimuthal dependence of the GISAXS intensity \cite{kaganer21gisaxs}.

\begin{figure}
\includegraphics[width=1\columnwidth]{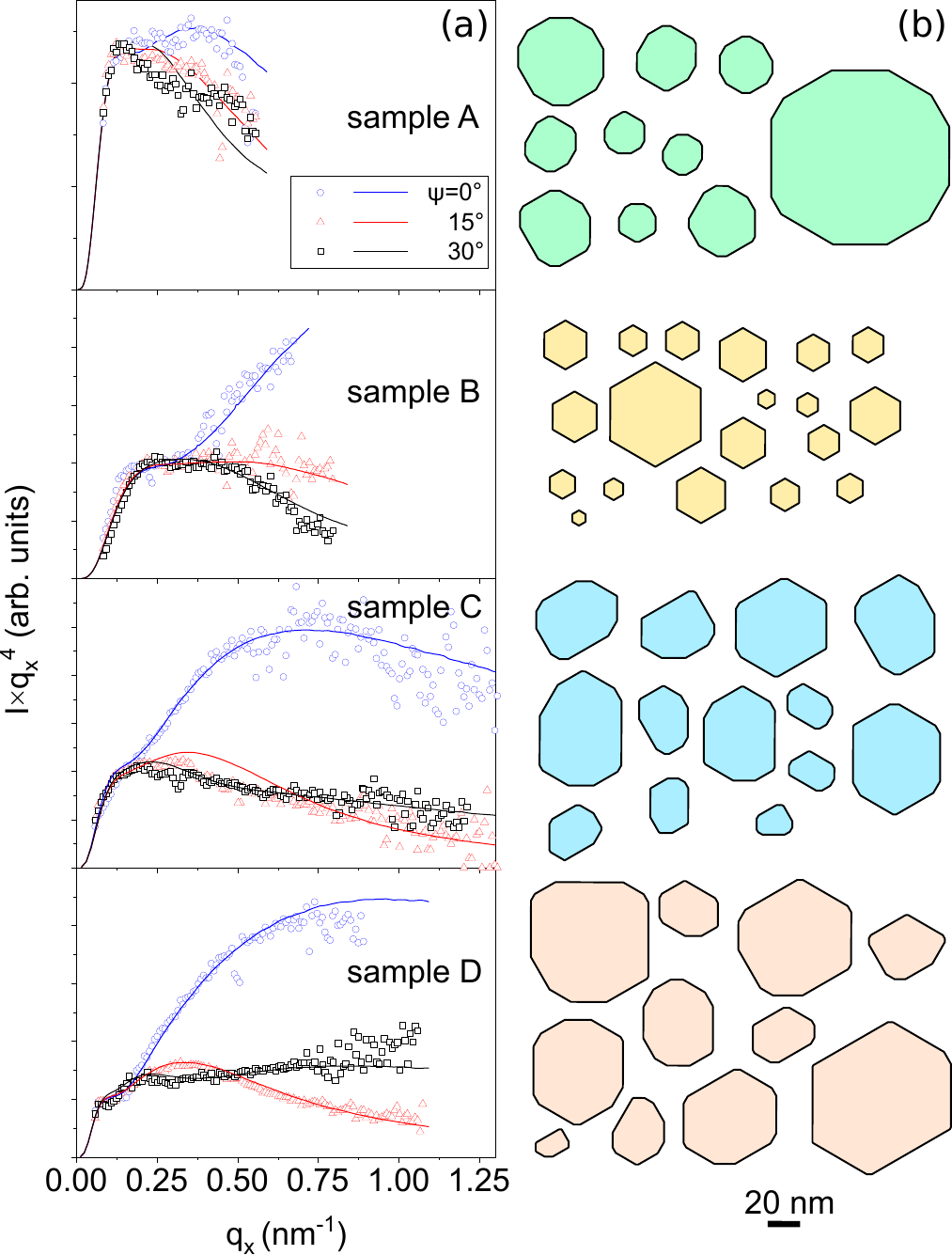}

\caption{(a) GISAXS intensities $I(q_{x})$ for samples A--D presented as
products $I(q_{x})q_{x}^{4}$ (open symbols) and respective Monte
Carlo simulations (lines). The measurements are performed for three
different azimuthal orientations $\psi$ of the incident X-ray beam
with respect to the side facets of NWs. (b) Examples of the cross
sections of NWs used in the simulation of each sample. The scale bar
is common for all samples.}

\label{fig:Iq4} 
\end{figure}

Hence, we plot in Fig.~\ref{fig:Iq4} the GISAXS intensity for samples
A--D as the product $I(q_{x})q_{x}^{4}$. The reciprocal space maps,
similar to the ones presented in Fig.~\ref{fig:maps}, were measured
with the azimuthal rotation of the samples about the vertical axis
on an angle $\psi$ from $0^{\circ}$ to $90^{\circ}$ with a step
of $15^{\circ}$. The sixfold symmetry of the scattering patterns
is established, and the distinct intensity distributions obtained
for $\psi=0^{\circ}$, $15^{\circ}$, and $30^{\circ}$ are shown
by circles, triangles, and squares respectively.

We model the GISAXS intensity by the Monte Carlo method, as described
in Ref.~\cite{kaganer21gisaxs}. The NWs are considered as prisms
with polygonal cross sections. Then, the scattered intensity from
a single NW can be expressed through the coordinates of the vertexes
of the polygon. The simplest shape of the cross section is the a regular
hexagon limited by GaN($1\bar{1}00$) facets. The hexagon sizes are
assumed to obey a lognormal distribution in the Monte Carlo simulations.
The calculated intensity curves in Fig.\ \ref{fig:Iq4} are obtained by averaging 
the scattering intensity over about $5\times10^5$ NWs generated on random, 
a number that is an order of magnitude smaller compared to the number of 
NWs contributing to the measured GISAXS intensity, estimated in Sec.\ \ref{sec:Experiment}.

In most cases, regular hexagons are not suitable to model the experimental
data. Thus, we first allow a random distortion of the hexagon while
keeping the GaN($1\bar{1}00$) side facets. Secondly, the vortexes
of the hexagons are cut on random by lines with the orientation average
of these of the adjoining sides, thus representing GaN($11\bar{2}0$)
side facets. In this case, the cross-sectional shape of the prism
becomes a dodecagon (a polygon with twelve vortices), rather than
a hexagon (a polygon with six vortices). The scattered intensity still
can be calculated using the positions of the vortices. Lognormal distributions
are assumed for the hexagon distortion and for the cuts of its apexes,
with the parameters varied to fit the experimental curves. For each
simulated polygon, we calculate its area $A$, perimeter $P$, as
well as the parts of the perimeter $p_{1\bar{1}00}$ and $p_{11\bar{2}0}$
representing separately ($1\bar{1}00$) and ($11\bar{2}0$) facets,
$p_{1\bar{1}00}+p_{11\bar{2}0}=P$. The quantities of interest are
the NW radius $R=2A/P$ and the fraction of ($11\bar{2}0$) facets
$f_{11\bar{2}0}=p_{11\bar{2}0}/P$.

The NW array is simulated by randomly rotating the NWs about the horizontal
and vertical axes in angular ranges of the tilt and twist determined
by the XRD measurements in Sec.~\ref{subsec:XRD} and presented in
Table~\ref{Table}. We also take into account roughness of the side
facets of NWs by including random shifts of the side facets in direction
of their normals by monolayer height steps. A geometric distribution
of steps is assumed and a roughness factor is calculated in the same
way as it is done in crystal truncation rod calculations \cite{Robinson1986a}.
As a result, the contribution of each facet to the scattering amplitude
contains an additional roughness factor \cite{kaganer21gisaxs}.

The lines in the plots in Fig.~\ref{fig:Iq4}(a) are obtained by
the Monte Carlo simulation, examples of the simulated cross-sectional
shapes are shown in Fig.~\ref{fig:Iq4}(b), and the parameters of
the NW ensembles of the samples A--D are included in Table~\ref{Table}.
Let us consider first the GISAXS intensity from sample B. Its dependence
on the azimuth $\psi$ is qualitatively similar to that of GaN NWs
on Si(111) \cite{kaganer21gisaxs}. The product $I(q_{x})q_{x}^{4}$
rises up at large $q_{x}$ at $\psi=0$, decays at $\psi=30^{\circ}$,
and shows an intermediate behavior at $\psi=15^{\circ}$. Such behavior
is expected for hexagonal cross sections, and we find that the simulation
of NWs by regular hexagons adequately describes the experiment. The
maximum intensity at large $q_{x}$ is the facet truncation rod scattering,
and the minimum at $\psi=30^{\circ}$ is the scattering in the direction
of the angle between the facets. The rise of the $I(q_{x})q_{x}^{4}$
curve at small $q_{x}$ is sensitive to the average NW radius, while
a dip before further rise of the curve at $\psi=0$ provides the width
of the radial distribution.

\begin{figure}[b]
\includegraphics[width=1\columnwidth]{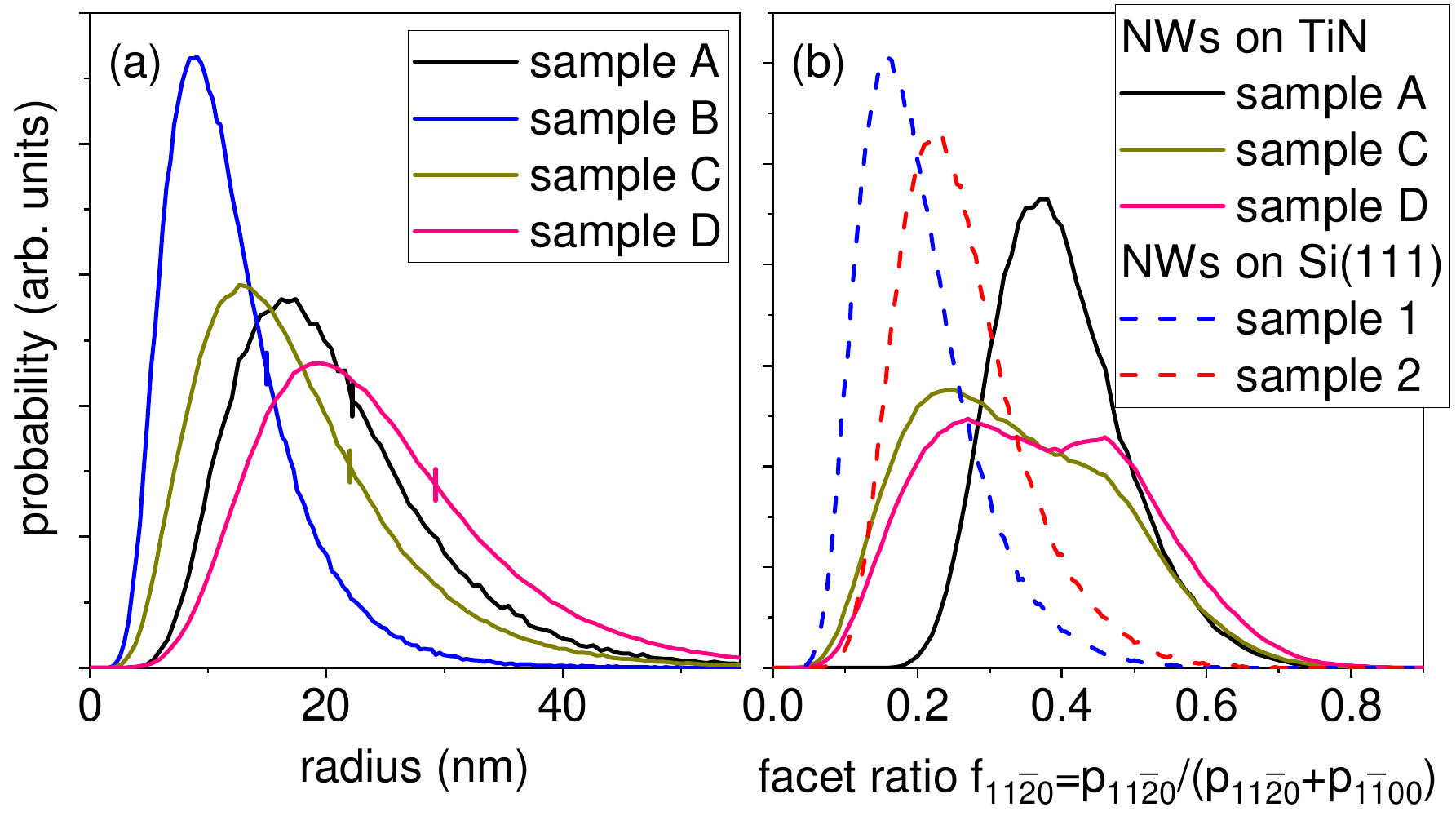}
\caption{Probability distributions of (a) radius and (b) fraction of the $(11\bar{2}0)$
facets obtained by the Monte Carlo simulations of the GISAXS intensity
curves in Fig.~\ref{fig:Iq4}. Vertical bars on the curves in (a) show the average
NW radii obtained in Ref.\ \cite{calabrese20} from SEM micrographs.}
\label{fig:RadiusPerim} 
\end{figure}

The distribution of the NW radii obtained in the Monte Carlo simulation
is shown in Fig.~\ref{fig:RadiusPerim}(a). We find a mean radius
of 11.2~nm and a standard deviation of the radial distribution of
5.2~nm. Sample B exhibits the thinnest NWs from the series under
investigation. We also mark by vertical bars on the respective curves in 
Fig.~\ref{fig:RadiusPerim}(a) the mean radii obtained in Ref.\ \cite{calabrese20}
from top view SEM micrographs and reproduced in Table \ref{Table}.
The mean radii thus obtained are systematically larger by 2 to 5~nm than 
the ones obtained from GISAXS in the present work. The difference can be
partially attributed to the limited resolution of SEM, and partially
to asymmetric radii distributions in Fig.~\ref{fig:RadiusPerim}(a),
with the maximum value smaller than the average. 

A fine tuning of the Monte Carlo simulations to the
experiment also requires to include roughness with an rms value of
$\sigma=0.7$~nm. Since the height of a monolayer step at the side
facet of a NW is equal to the GaN lattice parameter $a=0.319$~nm,
this value corresponds to only two atomic steps along the entire NW
length.

Proceeding now to sample C, we find that the intensity curve at the
intermediate azimuth $\psi=15^{\circ}$ is not in between the curves
for $0^{\circ}$ and $30^{\circ}$, as it is expected for oriented
hexagons and observed for sample B. Rather, the curves at $15^{\circ}$
and $30^{\circ}$ almost coincide, and at large $q_{x}$ the intensity
at the azimuth of $15^{\circ}$ is even smaller. Simulation of these
intensity curves requires to take into account both $(1\bar{1}00)$
and $(11\bar{2}0)$ side facets. Figure \ref{fig:Iq4} shows the result
of the Monte Carlo simulation for sample C and examples of the cross
sections used in the simulation. The NW radii are larger, and the
radii distribution is broader compared to sample B [see Fig.~\ref{fig:RadiusPerim}(a)].
The distribution of the fraction $f_{11\bar{2}0}$ is shown in Fig.~\ref{fig:RadiusPerim}(b).
A wide distribution of the ratios of different facets is needed to
model the experimental curves. For a comparison, dashed lines in Fig.~\ref{fig:RadiusPerim}(b)
show the facet fractions $f_{11\bar{2}0}$ obtained in the modeling
of the GaN NWs on Si(111) \cite{volkov22}. For NWs on Si(111), the
$(11\bar{2}0)$ facets are present in much smaller amounts.

Turning now to sample D, one can see in Fig.~\ref{fig:Iq4} that
the curves at the sample orientations $\psi=15^{\circ}$ and $30^{\circ}$
are swapped in comparison to sample B: the intensity at the intermediate
orientation $\psi=15^{\circ}$ (red triangles) is minimum and not
intermediate as it is for sample B. A good agreement of the experimental
curves and the Monte Carlo modeling of the intensity curves for sample
D is reached only with an even broader variation of the facet ratio,
see Fig.~\ref{fig:RadiusPerim}(b). A double-humped distribution
of the facet ratio for this sample seems an artifact of the modeling
of the cross sections by first producing hexagons and then cutting
the corners to obtain dodecagons. The NW radii for sample D are largest
in the series {[}see Fig.~\ref{fig:RadiusPerim}(a){]}. Comparing
the average radii and the widths of their distributions in Table \ref{Table},
one can see that the relative widths of the distributions (the ratios
of the standard deviation to the mean value) for samples B--D are
close.

Comparing now the intensity curves for sample A in Fig.~\ref{fig:Iq4}
with these for samples B--D, one can see that the difference between
the curves for different sample orientations $\psi$ is notably weaker,
and that the product $Iq_{x}^{4}$ decreases at large $q_{x}$ significantly
faster. A weak orientation dependence implies roundish NW shapes,
and such shapes are obtained in the Monte Carlo modeling, see Fig.~\ref{fig:Iq4}(b).
A fast intensity decay is a consequence of a large roughness of the
side facets. The modeling gives the roughness $\sigma=2.3$~nm, notably
larger than those for samples B--D (see Table \ref{Table}).

\section{Discussion}

GaN NWs grown on 3.4 \textmu m thick Ti films sputtered on Al$_{2}$O$_{3}(0001)$ possess remarkably small misorientation ranges, less than 2$^{\circ}$ out-of-plane (tilt) and less than 1$^{\circ}$ in-plane (twist), see Table \ref{Table}. For comparison, GaN NWs grown on the most common substrate Si(111) exhibit 3--5$^{\circ}$ tilt and twist \cite{jenichen11,geelhaar11,kaganer21gisaxs} because of the formation of an amorphous SiN$_x$ film, just a few nanometers in thickness, on the Si surface. The other extreme is the epitaxial growth of GaN NWs on AlN/6H-SiC$(000\bar1)$ with a tilt and twist of 0.4$^{\circ}$ and 0.6$^{\circ}$, respectively \cite{garrido14}.

The XRD reciprocal space maps in Fig.~\ref{fig:XRD}(a,b) show that,
as a result of the interfacial reactions between Ti and Al$_{2}$O$_{3},$
the sputtered homogeneous Ti film transforms into a heterogeneous
alloy containing topotaxial crystallites of Ti, Ti$_{3}$Al, and Ti$_{3}$O.
These crystals possess hexagonal symmetry, (0001) orientation, and
lattice parameters intermediate between those of Al$_{2}$O$_{3}$
and GaN. They are topotaxially oriented with respect to the substrate
with a misorientation of less than 1$^{\circ}$. The cubic TiO$_x$N$_y$ 
oxynitrides are also found in the film. Simultaneously with
the reaction of Ti with the Al$_{2}$O$_{3}$ substrate, the top 40--80~nm
of the Ti film are converted to cubic TiN, on which the NW growth
takes place \cite{calabrese19}. 

The GISAXS measurements and their Monte Carlo modeling allow us to
determine the distributions of the NW radii and their cross-sectional
shapes, as well as the roughness of their side facets. NWs of sample
A, grown on a 1.3~\textmu m thick Ti film, possess roundish cross-sectional
shapes and a relatively large roughness of the side facets. Sample A also exhibits blueshifted and broadened photoluminescence spectra as compared to reference GaN NWs, which was attributed to the incorporation of O atoms (diffusing from the Al$_2$O$_3$ substrate at the employed GaN growth temperature), resulting in a background doping high enough to screen excitons and induce bandgap renormalization and band-filling \cite{calabrese19}. We thus speculate that the presence of O adatoms at the NW sidewalls may modify their surface energy, giving rise to the presence of both $\langle 1\bar{1}00 \rangle$ and $\langle 11\bar{2}0 \rangle$ facets. To reduce this interdiffusion, further samples were grown
on 3.4~\textmu m thick Ti films \cite{calabrese19}.

We find that the roughness of the side NW facets of samples B--D
is at least twice smaller than that of sample A. However, the XRD
measurements do not show a notable difference in the reciprocal space
maps between the samples. The XRD peaks from Ti$_{3}$O remain the
most intense ones both in the symmetric reflection in Fig.~\ref{fig:XRD}(a)
and in the grazing incidence reflection in Fig.~\ref{fig:XRD}(b).
We note that the latter measurement reveals the structure of the top
part of the layer, due to small incidence and exit angles close
to the critical angle of total external reflection. Hence, O still diffuses
through the whole Ti layer despite its increased thickness. Regarding the NW shape, only sample B (even grown at the highest temperature) exhibits purely hexagonal NW cross sections, while samples C and D are again characterized by the coexistence of $\langle 1\bar{1}00 \rangle$ and $\langle 11\bar{2}0 \rangle$ facets. Furthermore, samples B--D display narrow excitonic transitions in their photoluminescence spectra \cite{calabrese20}, ruling out the incorporation of O exceeding a concentration of at most $10^{17}$~cm$^{-3}$. Our speculation above that O adatoms at the NW sidewalls may modify their surface energy is clearly not supported by these experimental facts.

A recent transmission electron microscopy study of GaN NWs on Si(111)
revealed a transformation of the cross-sectional shapes of NWs during
their growth \cite{volkov22}. The NWs possess hexagonal cross sections
with $(1\bar{1}00)$ side facets in their top parts. During growth,
their bottom parts attain roundish shapes with both $(1\bar{1}00)$
and $(11\bar{2}0)$ side facets present. The hexagonal shapes in the
top parts are considered as the equilibrium growth shape under the
impinging Ga and N fluxes, while the roundish shapes in the bottom
parts are understood to reflect a transformation to the equilibrium
crystal shape when the side NW surface is shadowed from the impinging
fluxes. Since the density of the GaN NWs on TiN studied in the present work
is at least one order of magnitude lower compared to GaN NWs on Si(111),
the NWs receive impinging fluxes along their whole length, i.\,e., the mechanism considered for GaN NWs on Si(111) cannot be invoked to explain the roundish shape of the NWs and the presence of $\langle 11\bar{2}0 \rangle$ facets.
 
A radically different possibility arises from the incorporation of substantial amounts of Ga prior to GaN growth into the Ti layer \cite{calabrese19}. After the growth and during cooling, some of the Ga may be released from the Ti layer due to the reduced solubility at lower temperatures. This Ga wets the GaN NWs and turns into GaO$_x$ upon air exposure. We note that the GISAXS intensity depends only on the density of the matter and does not depend on its crystallinity. Since crystalline GaN and amorphous GaO$_x$ have close densities, the cross-sectional shapes obtained in Fig.\ \ref{fig:Iq4}(b) are the ones of the NWs covered with the GaO$_x$ shell, if the latter is present. The roughness obtained from the GISAXS study also applies the outer NW surface. The roundish shape and the coexistence of $\langle 1\bar{1}00 \rangle$ and $\langle 11\bar{2}0 \rangle$ facets would thus be a characteristics of the GaO$_x$ shell, while the GaN core may very well have regular hexagonal shape. Plan-view transmission electron microscopy and electron dispersive x-ray spectroscopy could be used to refute or confirm this hypothesis. 

\section{Summary}

The diffusion of Al and O from the Al$_2$O$_3$ substrate to the sputtered Ti film gives rise to topotaxial crystallites of Ti, Ti$_3$Al, and Ti$_3$O possessing very little misorientation with respect to the substrate. GaN NWs grown on this film are epitaxially oriented with respect to the substrate notably better compared to GaN NWs on Si(111).  

The GISAXS intensity together with its Monte Carlo modeling is capable to provide detailed information on the NW arrays, particularly the distributions of the cross-sectional sizes of the NWs, the fractions of the $(11\bar20)$ and $(1\bar100)$ side facets, and the roughness of these facets. The NW radii obtained from GISAXS are systematically smaller by 2 to 5 nm compared to the ones obtained from SEM micrographs. The fraction of the $(11\bar20)$ facets is notably larger compared to GaN NWs on Si(111), so that the NWs have roundish cross-sectional shape. An exclusion is the sample grown at the a highest temperature. The GISAXS intensity is highly sensitive to the roughness of the side facets, a parameter hardly accessible by any other method. We find that the roughness of the micron long side facets does not exceed the height of 2--3 atomic steps. We propose that both the roughness and the shape are the result of the presence of Ga adatoms at the NW sidewall after growth, and the formation of a GaO$_x$ shell upon exposure of the NW to the ambient.

\begin{acknowledgments}
The authors thank R. Volkov and N. Borgardt for fruitful discussions, Thomas Auzelle
for critical reading of the manuscript, and the ESRF for the provision
of beam time. S. F.-G. acknowledges the partial financial support received through the Spanish program Ramón y Cajal (co-financed by the European Social Fund) under grant RYC-2016-19509 from Ministerio de Ciencia, Innovación y Universidades. He also thanks Universidad Autónoma de Madrid for the transfer to Universidad Politécnica de Madrid of part of the materials and equipment purchased with charge to the RYC -2016-19509 grant.
\end{acknowledgments}


%

\end{document}